\documentclass[conference]{IEEEtran}

\usepackage{subcaption}

\usepackage{soul}
\usepackage{color}

\ifCLASSINFOpdf
  \usepackage[pdftex]{graphicx}
\else
   \usepackage[dvips]{graphicx}
\fi

\usepackage{algorithm}
\usepackage{algpseudocode}

\usepackage{blindtext}
\let\labelindent\relax 
\usepackage{enumitem}

\begin{document}
%
\title{Big Holes in Big Data: \\ A Monte Carlo Algorithm for Detecting Large Hyper-rectangles in High Dimensional Data}

\author{\IEEEauthorblockN{Joseph Lemley}
\IEEEauthorblockA{Computer Science Department\\
Central Washington University\\
Ellensburg, USA\\
Email: lemleyj@cwu.edu}
\and
\IEEEauthorblockN{Filip Jagodzinski}
\IEEEauthorblockA{Computer Science Department\\
Western Washington University\\
Bellingham, USA\\
Email: filip.jagodzinski@wwu.edu }
\and
\IEEEauthorblockN{R\u azvan Andonie}
\IEEEauthorblockA{Computer Science Department\\ Central Washington University\\Ellensburg, USA\\ and\\  Electronics and Computers Department\\ Transilvania University of Bra\c{s}ov, Romania \\
Email: andonie@cwu.edu}}


%


\maketitle

\begin{abstract}
We present the first algorithm for finding holes in high dimensional data that runs in polynomial time with respect to the number of dimensions.  Previous algorithms are exponential. Finding large empty rectangles or boxes in a set of points in 2D and 3D space has been well studied. Efficient algorithms exist to identify the empty regions in these low-dimensional spaces. Unfortunately such efficiency is lacking in higher dimensions where the problem has been shown to be NP-complete when the dimensions are included in the input. Applications for algorithms that find large empty spaces include big data analysis, recommender systems, automated knowledge discovery, and query optimization. Our Monte Carlo-based algorithm discovers interesting maximal empty hyper-rectangles in cases where dimensionality and input size would otherwise make analysis impractical. The run-time is polynomial in the size of the input and the number of dimensions. We apply the algorithm on a 39-dimensional data set for protein structures and discover interesting properties that we think could not be inferred otherwise. 

\end{abstract}


%
\IEEEpeerreviewmaketitle

\section {Introduction}
An important aspect of data mining research is discovering large empty areas (or holes) in data sets. If we view each item (or tuple) in a data set as a point in a \textit{k}-dimensional space, then a hole is a region in the space that contains no data points. In a continuous space, there exist a large number of holes because it is not possible to fill up the continuous space with data points. The existence of large holes is mainly due to the following two reasons:

\noindent
\begin{enumerate}
\setlength{\leftmargin}{-2in}
\item The data cases collected are insufficient, resulting in some regions having no data point.
\item Certain value combinations are not possible. For example, assume we have a  data set with two continuous attributes X and Y. Both X and Y can take values from 1 to 10, but it is observed that {\tt if X $>$ 5, then Y $<$ 4}. In other words, there exists an empty area, i.e., a rectangular region defined by {\tt 5 $<$ X $\le$ 10 and 4 $\le$ Y $\le$ 10}.
\end{enumerate}
Organizations often deploy systems to generate rules about values that are present in their data. However these rules often do not provide the user with complete information about their data set. For example, a particular organization may rely on a learning system to generate a set of rules from their database. One of the rules may be the following:
\begin{center}
\texttt{ if Company Size $>$ 600 then Service = Yes}
\end{center}

\noindent
This rule specifies that if the company has more than 600 employees, then the company uses the service provided by the organization. Now assume the company size is partitioned into 10 categories. A close inspection of the database may reveal that no company, whose size is in the range of 400-800, uses the service. Hence, there is a hole in the data. Realizing the existence of this hole may lead the organization to probe into the possibilities of modifying its service or of doing more promotion to attract the medium size companies.

In the case of a one-dimensional data set, finding the largest hole becomes the problem of finding large gaps and is trivial to solve. When the number of dimensions increases, the problem quickly becomes complex and has been shown to be NP-hard when the number of dimensions is included in the input. In two or more dimensions, locating large holes is very difficult. Consider for example Table~\ref{tbl:Sample}, which lists 4 entries in a hypothetical 7-dimensional data set. Locating the largest hole in the seven-dimensional space is very difficult.

\vspace{-0.3cm}
\begin{table}[h]
  \caption{Sample 7-dimensional data. Finding the largest 7-dimensional hole is not intuitive.}
  \centering
  \begin{tabular}{| c | c | c | c | c | c | c |}
    \hline
    \textbf{ID} & \textbf{Age} & \textbf{GPA} & \textbf{Gender} & \textbf{Height} & \textbf{Weight} & \textbf{Income}\\
    \hline
    1 & 20 & 3.6 & male & 60in. & 170lb & \$100,000\\
    \hline
    2 & 19 & 4.0 & female & 75in. & 160lb & \$20,000\\
    \hline
    3 & 21 & 3.7 & female & 71in. & 250lb & \$94,000\\
    \hline
    4 & 26 & 3.4 & female & 62in. & 150lb & \$112,000\\
    \hline
  \end{tabular}
  \label{tbl:Sample} 
\end{table}

In general, a data set contains a large number of holes because each item is only a point in a \textit{k}-dimensional space. Even if each attribute takes discrete or nominal values, it may be still quite difficult to fill the entire space. However not all holes are interesting. Actually, most of them are not. We concur with Liu \textit{et al.} \cite{liu1997discovering} regarding the types of holes that tend to be interesting and uninteresting.

\noindent
The following types of holes are not interesting \cite{liu1997discovering}:

\begin{enumerate}
\item Small holes: they exist because there are only a limited number of cases in the data set.
\item Known impossible value combinations: they exist because certain value combinations are not possible and this fact is known previously.
\end{enumerate}

\noindent
Certain types of holes can be of great importance:

\begin{enumerate}
\item Holes that represent impossible value combinations that are not known previously.
\item Holes that indicate that the data collected within those areas are insufficient.
\item Holes that are suspected by the user and need confirmation.
\end{enumerate}

Holes can take the form of any shape. For simplification, we focus particularly on the problem of axis-aligned rectangular holes in arbitrarily high dimensions. Holes of this type are easier to understand \cite{liu1997discovering}:  (especially in high dimensions) and have the useful property that they can be easily converted to if/then rules. Formally, we define a Maximal Empty Hyper-Rectangle (MEHR) as follows:
\par
\begingroup
\leftskip2em
\rightskip\leftskip
\noindent
\textbf{Given a \textit{k}-dimensional space \textbf{S} containing \textit{n} points, where each dimension is bounded by a minimum and maximum value, a  MEHR  is a hyper-rectangle that has no data point within its interior and has at least one data point on each of its 2\textit{k} bounding surfaces. }
\par
\endgroup

\begin{figure}[!htb]
\begin{centering}
  \includegraphics[width=0.45\linewidth]{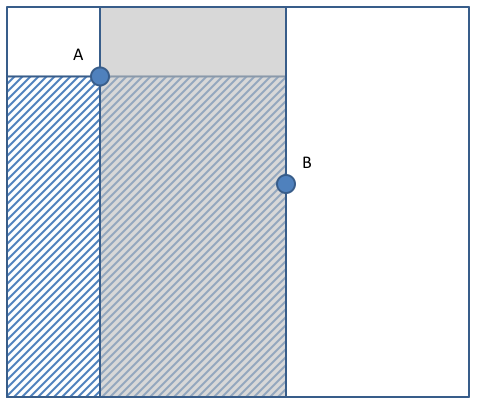}
  \caption{Two sample points, A and B, in 2D space. Both of these points are bounding points for the blue thatched and gray rectangles, that overlap. The top and bottom edges of the gray rectangle are bound by the minimum and maximum values of the \textit{y}-direction, while the left and bottom edges of the blue thatched rectangle are bounded by the minimum \textit{y} value and minimum \textit{x} values of the 2D space.}
  \label{fig:Bounding}
\end{centering}
\end{figure}

\noindent
We call these points the bounding points of the hyper-rectangle. Each side \textit{i} of the MEHR is parallel to one axis of \textbf{S} and orthogonal to all the others. (See Figure~\ref{fig:Bounding}).

We are interested in those MEHRs that are sufficiently large or significant. The user can specify how to measure the size of a MEHR and what size is considered sufficiently large. These are all application dependent. A simple way of measuring the size of a MEHR is by its volume. We follow the convention used by ~\cite{ordonez1999clustering} \cite{dumitrescu2014computational} \cite{backer2010mono}  of a generalized definition of volume (or hyper-volume) when speaking of the size of rectangular objects in arbitrary dimensions. For axis aligned hyper-rectangles, this is the product of the distance between the smallest and largest points on each axis which indicate the boundary points of this shape. 

Our objective is to find all the MEHRs in the user-specified \textit{k}-dimensional space that satisfy the sufficiently large criterion and to rank them according to their sizes.

The upper bound on the number of MEHRs has been found to be in $O(n^{k-1})$ \cite{liu1997discovering}, where $n$ is the number of $k$-dimensional data points. Finding all MEHRs has been shown to be NP-hard~\cite{eckstein2002maximum}. Previously developed approaches are unsuitable for large data sets primarily because they must locate and rank all MEHRs to find the largest MEHR. 

This gave us the motivation for our work.We present the first efficient, polynomial time, algorithm for discovering large interesting hyper-rectangles in a \textit{k}-dimensional space. The  algorithm runs in polynomial time with respect to \textit{k}.  To the best of our knowledge, all previous algorithms designed to solve this problem optimally are exponential. Our solution is a Monte Carlo approach which does not necessarily find the optimal solution. In experiments, our algorithm identifies the same large holes as existing algorithms for low dimensions (where all existing algorithms perform well) but is also able to quickly identify large holes in high dimensional spaces that would not be feasible with existing approaches that exhaustively identify all of the rectangles and then find the largest from among them. Our algorithm is well suited for large data sets with many attributes. 

We demonstrate the speed and accuracy of our approach by comparing the time taken to find the largest hole using our method compared with the method used in \cite{liu1997discovering}.

After experimentally demonstrating that our algorithm produces correct results orders of magnitude faster than the comparison method on publicly available machine learning datasets, we use our algorithm on a larger 39-dimensional data set. We show how to use information about large empty regions to infer interesting relationships among dimensions that would have been difficult to discover using other techniques.

Section \ref{relatedwork} describes related work. In Section \ref{algorithm} we introduce our Monte Carlo algorithm. Section \ref{validate} contains the results of experiments used to assess the efficiency of our approach. Section \ref{application} illustrates the use of our algorithm on a real-world application in bioinformatics. We conclude with final remarks and open problems in Section \ref{conclusions}.

\section{Related Work} \label{relatedwork}

The problem of finding the maximum empty rectangle in a set of points has been studied in depth in low dimensions and has been proven intractable in high dimensions~\cite{backer2009bichromatic} \cite{eckstein2002maximum}. 

In one dimension, the problem of finding the largest maximum space simplifies to the Maximum Gap problem \cite {preparata1985computational} for which there exists a linear time solution based on the pigeonhole and bucketing principles. Preparata \textit{et al.} suggest that no generalization of this approach seems possible in higher dimensions \cite {preparata1985computational}.

In two dimensions, the problem of finding the largest hole becomes the ``largest rectangle''  problem.  One of the fastest solutions was proposed in \cite{aggarwal1987fast} and runs in $O(n\,log^2\,n)$. 

An efficient algorithm for finding the largest empty hyper-rectangle in 3D space also exists, developed by Datta, \textit{et al.}~\cite{datta2000efficient}. Their algorithm runs in $O(n^3)$  and they report an average case complexity of $O(n\,log^4\, n)$. 

The first algorithm to identify the largest maximal empty hyper-rectangle in an arbitrary number of dimensions was introduced in \cite{liu1997discovering} and \cite{ku1997discovering}. The algorithm runs in $O(n^{2k-1}\, k^3\,(log\,n)^2)$ time and $O(n^{2k - 1})$ space. It relies on a heuristic \textit{BigEnough()} which allows small rectangles to not be considered. This heuristic substantially narrows the search space and improves execution time. In the context of large data sets, a main disadvantage of this algorithm is that it requires calculating and storing every empty hyper-rectangle larger than \textit{BigEnough()} in memory, and that it requires that every element be processed before anything can be known about the size of large holes. Our implementation of \cite{liu1997discovering} performs well with small data sets (fewer than 1000 entries with 5 or fewer dimensions), and we recommend their algorithm for small to mid-sized data sets when one can reasonably estimate \textit{BigEnough()}.

Realizing the limits of this approach, Liu \textit{et al.} \cite{liu1998using} introduced a method based on decision tree induction to facilitate discovery of large and interesting hyper-rectangles. 
 
Edmonds \textit{et al.} \cite{edmonds2001mining, edmonds2003mining} utilized a technique similar to Aggarwal and Suri's method \cite {aggarwal1987fast}, with a focus on database applications. Their algorithm for calculating maximal empty hyper-rectangles runs in $O(k\,n^{2k - 2})$ and has the space complexity in $O(k\,n^{k-1})$.

Eckstein proved that the maximal box problem (and thus the maximal empty rectangle problem) is NP-hard, by reduction from the maximal clique problem~\cite{eckstein2002maximum}.  

Another related approach involves a query point algorithms to find the largest maximal empty hyper-rectangle that contains only one query~\cite{gutierrez2012finding}. In this method, a quadtree is used to realize a significant speed improvement, and has the advantage that there is no need for all the found rectangle objects to be maintained in memory. The approach has been shown to be especially useful for query optimization in databases. 

Geometric approaches to join operations that use information about holes on databases have had much recent success as demonstrated by \cite{AboKhamis2015}.

A related problem, the bichromatic rectangle problem, deals with finding the largest empty rectangle that contains only blue but no red dots. Backer \textit{et al.} described a set of solutions to this problem, and showed that it is identical to the maximum empty  hyper-rectangle problem~\cite{backer2009bichromatic}. Furthermore, they present an exact algorithm running in $O(n^k\,log^{2k}\,n)$ time, for any \textit{k} $\ge3$.

In \cite{dumitrescu2009} Dumitrescu \textit{et al.} presented the first efficient $1-\epsilon$ approximation algorithm for the problem of finding the size of the largest MEHR. They showed that the minimum size of any maximum maximal empty hyper-rectangle has a volume of $1/n$ on a unit cube.

\section{Our Algorithm: A Monte-Carlo Approach} \label{algorithm}

In the following, we describe our algorithm, composed of four steps. 

\subsection{The Algorithm}

\textbf{Step 1:} Algorithm Preparation

As a first step, we retrieve points from a file or database, removing any attributes for which distance is undefined or attributes where each element has the same value.  Each remaining attribute column is assigned a number from 0 to \textit{k}-1, and is treated as a dimension in a \textit{k} dimensional space. Next, we scale the data to the range [0, 1]. For each dimension, we create a sorted list of attribute values with duplicates removed, which is important for a later step in being able to guarantee the emptiness of generated maximal empty hyper-rectangles.  These lists can be considered as orthogonal projections of all points onto an axis.

\textbf{Step 2:} Generating Maximal Empty Rectangles

\begin{figure}[!htb]
\begin{centering}
  \includegraphics[width=0.9\linewidth]{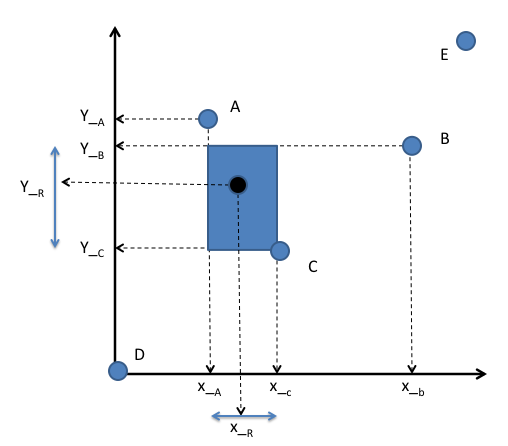}
  \caption{Generating Maximal Empty Rectangles. Five data points, A through E, in 2D. Points D and E constrain the maximal size of dimensions \textit{x} and \textit{y}. Points A through C are projected onto the \textit{x} and \textit{y} axes, shown as $y_a, y_b, y_c$ and $x_a, x_b, x_c$. A Monte Carlo randomly selected point, the black dot at coordinates $x_r, y_r$ is chosen. Its \textit{x} and \textit{y} coordinates are expanded (blue arrows) in both directions,  until another point in that coordinate is detected. The maximal expansion of both axes specifies the maximum bounding rectangle.}\label{fig:queryMethod}
\end{centering}
\end{figure}

In this phase, we create maximal empty hyper-rectangles using randomly generated query points that lie between any two points in our sorted list explained in \textbf{Step 1}, as shown in Figure~\ref{fig:queryMethod}. In 2D, the nearest points to the randomly selected point in the positive and negative directions of each dimension specify the bounding points of an initial empty hyper-rectangle along that dimension. Because these points are distinct and sorted, we know that the resulting rectangle is empty.

\noindent
The found rectangle is empty due to the following:

\begin{itemize}
\item For a rectangle to be non-empty there must exist at least one point inside it. 
\item A point can be said to be inside a rectangle if, for each dimension $k_i$, the value of the point at $k_i$ is between the upper and lower bounds of the rectangle at $k_i$. 
\item Assume the point is in the hyper-rectangle making the hyper-rectangle non empty. Then the value of a point at $k_i$ is in a sorted list of distinct points, which means that the point must be either the upper bound or lower bound of the rectangle (from description of how we create rectangles). This leads to a contradiction because the point cannot be both a bounding point and a point between bounding points without having a value greater than itself or less than itself.
\end{itemize}

\textbf{Step 3:} Expansion

Once a rectangle has been found, we want to enlarge it until it is maximal while still being empty. A hyper-rectangle is maximal and empty if it cannot be expanded along any dimension without containing one or more points. In Figure~\ref{fig:queryMethod} the found rectangle is not maximal because it can be expanded into several directions before it borders either the boundary of the space or is abutting one of the points.

We have developed several strategies for expanding rectangles that produce maximal empty hyper-rectangles, but care must be taken not to exclude or bias rectangles with regard to any particular dimension.  For example, if we always expand dimension 1 first, then dimension 2, and so on, then dimension \textit{k}-1 of our rectangle will usually have a much smaller width than it should have if we had processed them in random order. It is therefore important to randomize the order of expansion so that no single attribute is given any bias.

When expanding rectangles, there are 3 strategies that are useful depending on the features of the hyper-rectangles one is looking for.

In all expansion strategies we shuffle the order that dimensions are processed in. 

\label{section:expansion1}
\subsubsection{Expansion Approach 1}

\begin{figure}[!htb]
\begin{centering}
  \includegraphics[width=0.98\linewidth]{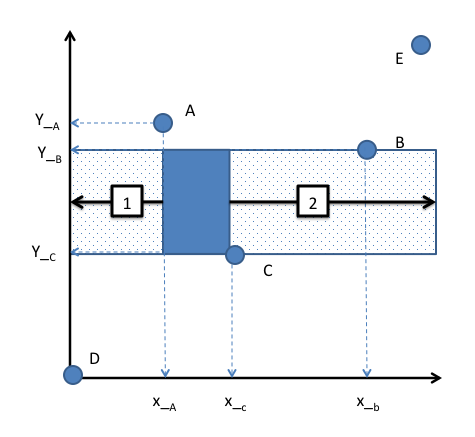}
  \caption{Expansion Approach 1 expanding maximally in a direction. Here, the \textit{x} axis was randomly selected to expand first, and it was expanded first into the negative direction, then the positive (shown numbered as black boxes), to generate the maximum rectangle indicated with blue dots.}
  \label{fig:expansion1}
\end{centering}
\end{figure}

For each dimension, maximally expand along that dimension before proceeding to the next. This strategy will find maximal rectangles that have many sides that are shared with the unit square, and it is particularly useful when we want to find ranges of values for which most attributes, if eliminated, have no meaningful impact on the resulting if/than rules. These rectangles will be maximally wide (bound by 0 and 1) in some dimensions, but more narrow in others, as shown in Figure~\ref{fig:expansion1}. 

\label{section:expansion2}
\subsubsection{Expansion Approach 2}

Expand along each dimension equally, stop expanding along a dimension when it becomes bound by a point and keep expanding along the other dimensions.  This strategy tends to find maximal hyper-rectangles for which the widths along each dimension are similar.

\begin{figure}[h]
    \centering
    \begin{subfigure}[b]{0.42\textwidth}
        \centering
        \includegraphics[width=\textwidth]{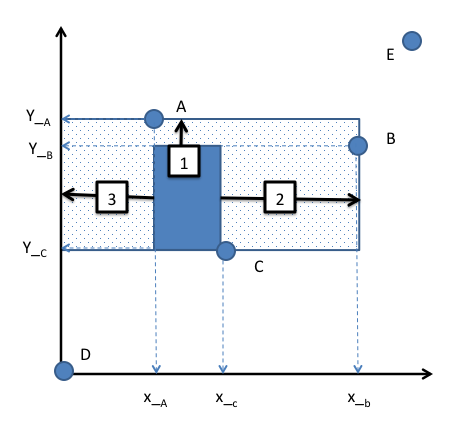}
        \caption{Expanding "up" first, then "right," and finally "left"}

    \end{subfigure}\\
    \begin{subfigure}[b]{0.42\textwidth}
        \centering
        \includegraphics[width=\textwidth]{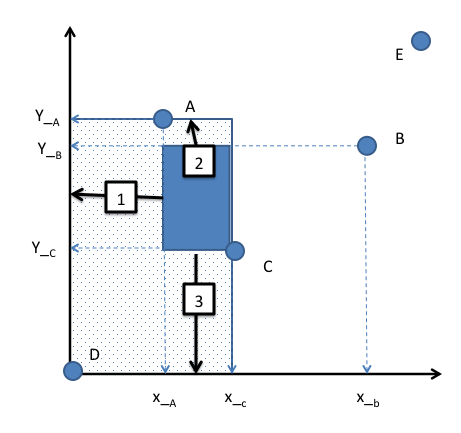}
        \caption{Expanding "left," then "up" the "down"}

    \end{subfigure}\\
    \caption{Expansion Approach 2. The order of how dimensions are expanded determines which hyper-rectangle is found. Numbers in squares specify the order of expansion. In (a), expanding in the "up" direction first, followed by the "right" direction, and finally to the "left" gives a different maximum hyper-rectangle than if the expansion order were different. Note that the second expansion in (a) prevents the eventual maximum rectangle from proceeding below point C, but in (b) the different order of expansion finds a maximum rectangle that continues to the the bottom-most boundary formed by point D.}
\end{figure}

\label{section:expansion3}
\subsubsection{Expansion Approach 3}

We expand a random amount along each dimension until the rectangle is maximally expanded in all dimensions without containing any points.  This strategy contains no statistical bias for any particular type of hyper-rectangle.  

\textbf{Step 4:} Finding interesting maximal empty hyper-rectangles

Our goal is to find interesting maximal empty hyper-rectangles that give insight into the data set. We propose a method to find empty hyper-rectangles that are potentially interesting. Like in \cite{liu1997discovering} and \cite {edmonds2001mining}, we assume that large empty rectangles, particularly rectangles with volumes close to the size of the largest maximal empty hyper-rectangle, have the greatest chances of being truly interesting. In order to exploit this, we take the approach of discounting any empty rectangle with a volume less than $\frac{1}{n}$.

\begin{figure}[!htb]
\begin{centering}
  \includegraphics[width=0.9\linewidth]{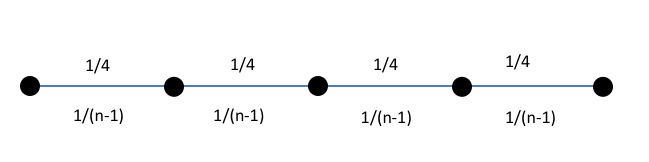}
  \caption{Five equidistant points create four maximal holes each with an area $\frac{1}{n-1}$. }
  \label{fig:1n}
\end{centering}
\end{figure}

This is reasonable because, according to Dumitrescu \emph{et al.} \cite{dumitrescu2013largest}, for a fixed k, the maximum volume is on the order of $ \Theta(\frac{1}{n})$. The volume of the largest box is $\Omega(\frac{1}{n})$.  Furthermore, unless the points are distributed in an equidistant manner, any large rectangle will have greater volume than $\frac{1}{n}$. For this reason, maximal empty hyper-rectangles with volumes of $\frac{1}{n}$ or less can be described as clusters, or areas with many points close together. Figure~\ref{fig:1n} shows that, in a 1D space, five equidistant points create four maximal holes each with areas proportional to approximately $\frac{1}{n}$. 

\begin{figure}[!htb]
\begin{centering}
  \includegraphics[width=0.8\linewidth]{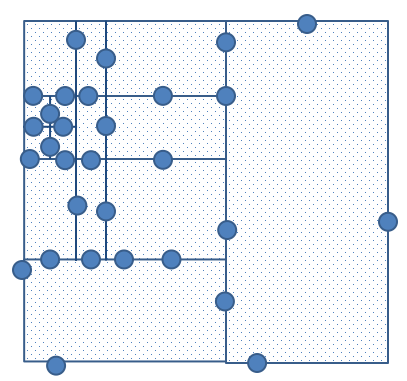}
  \caption{For a dart randomly thrown on a board with both small and large rectangles, the chances of it landing in a small rectangle are much less than landing in a large rectangle.}
  \label{fig:clusters}
\end{centering}
\end{figure}

Fortunately, due to the nature of our algorithm, we only very rarely find any rectangles with a volume less than $\frac{1}{n}$ because we have a strong statistical bias for finding large rectangles. To visualize this, imagine a dart randomly landing on a board with both small and large rectangles drawn on it. The chances of it landing in a small rectangle are much less than landing in a large rectangle, as shown in Figure~\ref{fig:clusters}.

This tends to become even more of a factor as the number of dimensions $k$ increases.  For example, in one of our computational experiments on our artificial data set, we landed in a rectangle with a volume less than or equal to $\frac{1}{n}$ only 14 times out of 108,509 randomly selected initial points.

Algorithms \ref{main} and \ref{create} provide the pseudocode of our approach.

\begin{algorithm}
\caption{Pseudocode}\label{main}
\begin{algorithmic}[1]
\Procedure{findMEHRs}{data, stop}\Comment{}
\State $data\gets normalize(data)$; \Comment{Normalize 0 to 1 in each dimension}
\State $projections\gets createOrthProjections(data)$;
\State $n\gets data.size()$
\State $tooSmall\gets \frac{1}{n}$
\State $c\gets 0$
\State $maxFound \gets 0$ 
\While{$c>stop$}\Comment{Run Monte Carlo step}

\State $mehr \gets createMEHR(data,projections)$
\If {$mehr.volume() > MaxFound$}
\State $maxFound \gets \Call{mehr.volume}$
\State \Call {mehrList.append}{mehr}
\State $c\gets 0$

\Else 
\If {$mehr.volume()>tooSmall$}
\State $c\gets c+1$
\EndIf
\EndIf
\EndWhile\label{mainendwhile}
\State $mehrList.sort()$
\EndProcedure
\end{algorithmic}
\end{algorithm}

\begin{algorithm}
\caption{createMEHR}\label{create}
\begin{algorithmic}[1]
\Procedure{createMEHR}{data, projections}
	\ForAll{$k \in projections$}
		\State $r\gets rand(0,1)$
		\State $L_k$,$U_k$ $\gets findBoundingPoints(r,projections)$\Comment{Binary Search to find bounds on k}
	\EndFor
	\State $ehr\gets createR(U,L)$ \Comment{Create MEHR from bounding points} 

	\State $mehr\gets expandRectangle(ehr,projections)$ \Comment{ Expand rectangle using one of the strategies discussed} 		
\Return $mehr$
\EndProcedure
\end{algorithmic}
\end{algorithm}

\subsection{Complexity}

Our algorithm's complexity is determined by the number of comparisons between the upper and lower bounds along each dimension with the value of a point on that dimension. 

To create an empty hyper-rectangle, we perform a binary search for the lower and upper bounds surrounding a randomly generated point. The number of comparisons is in $O(k\, log \,n)$ if the number of dimensions are included in the input.

Our enlargement step works by expanding the empty hyper-rectangle along each axis until it is no longer empty. The maximum number of expansions for each dimension is equal to the number of distinct points in the orthogonal projection of the unit hyper-cube onto that axis. This number is at most $n$. Therefore, the complexity of an expansion is  in $O(k\, n)$. It is important to note that this maximum can only be realized in extreme cases, such as when all points are bounding points of the unit square in all dimensions. 

Every time we expand we must verify that the rectangle is still empty, which requires $2n$ comparisons on $k$ dimensions meaning at most $2nk$ comparisons (the two comes from the fact that we must compare the upper and lower bounds of the rectangle with the value of the point).  This means our growth step is in $O(k^2\,n^2)$ with respect to the number of comparisons. 

Considering both processes of creating an initial empty hyper-rectangle and growing that rectangle, the overall complexity of our algorithm is in $O(k^2\,n^2 + k \,log\, n)$ = $O(k^2\, n^2)$. To the best of our knowledge, this is the first algorithm for finding  holes in data which has polynomial time with respect to the number of dimensions. 

Finally, the space complexity of our algorithm is in $O(n\, k)$ for storing the data points. This accomplishes our goal of an algorithm that scales well with the number of dimensions. 

\section{Validity} \label{validate}

It is necessary to not only demonstrate that our algorithm is fast but also that, despite its use of randomization, it produces the largest hyper-rectangle with regularity. To accomplish this goal we compare our algorithm's performance on various published machine learning benchmarks. In all cases we used the full data set first, and only reduced the size in cases where it was necessary to obtain a result from existing algorithms for comparisons. 

Liu's algorithm \cite{liu1997discovering} was chosen for the purpose of comparison because it has been proven to identify the largest empty hyper-rectangle and its \emph{BigEnough()} heuristic greatly cut down on computation time compared to other solutions. It is also the fastest published algorithm available for comparison. We measure the amount of time required for our algorithm to find the same rectangle produced by \cite{liu1997discovering}. 

The goal of these experiments is to show that, despite starting with random points, in all our experiments, our algorithm consistently finds the same largest empty hyper-rectangle in a reasonable amount of time. By ``consistent'' we understand that even if we cannot guarantee that the algorithm will always find the best solution, practically, this is usually the case. 

To facilitate reproducibility we use four data sets from the well known UCI machine learning repository \cite{Lichman:2013}.
\begin{enumerate}

\item The Iris data set contains measurements of the septal and petal widths and lengths for 3 classes of iris plants with 150 items total and 4 dimensions. 
\item The Combined Cycle power plant data set \cite{tufekci2014prediction} contains 9568 attributes and 5 dimensions. The dimensions are composed of physical measurements such as temperature, humidity, pressure, exhaust vacuum, and electric output as measured by sensors around the plant. 
\item The User Knowledge modeling data set contains information about student knowledge of electrical DC machines. It has 5 dimensions and 255 items. We only use the first 100 points of this data set because the comparison algorithm was unable to finish within a reasonable amount of time on the full 255 points. 
\item The Wilt dataset \cite{johnson2013hybrid} is a high-resolution remote sensing data set containing information about tree wilt. We use the training set without the categorical data (i.e., 4339 5-dimensional vectors). 

\end{enumerate}

\begin{table}[htb]
\centering
\caption{Execution Time}
\label{validity-table}
\begin{tabular}{|l|l|l|}
\hline
data set & Liu's algorithm  (s) & Our algorithm (s)\\ \hline
1        & 3.845       & 0.077   \\ \hline
2        & 521.426     & 97.46     \\ \hline
3        & 1523.75     & 0.44      \\ \hline
4        & 19916.4    & 3.6      \\ \hline 
\end{tabular}
\end{table}

The above table shows the runtime of our implementation of  \cite{liu1997discovering} compared with our algorithm using expansion strategy 3 averaged over 100 runs. The databases we chose represent a variety of types of data.  Additionally, we found the same largest MEHR but in a fraction of the time, which is an added benefit of our approach.

\section{An Application} \label{application}

To demonstrate the use of our algorithm, we apply it a large data set of 39 metrics for approximately 5,000 protein structures. We will show how the  algorithm's output  can provide insights about the data that previously were not known.

\subsection{Motivation and Data Set Description}
Proteins are three-dimensional dynamic structures that mediate virtually all cellular biological events. They are composed of long chains of amino acids, and knowing how they move in order to perform their functions is fundamental in designing medicines that regulate disease-causing proteins.

The Protein Data Bank (PDB) is a repository of the atomic $x,y$ and $z$ coordinates of over 100,000 protein structures that have been solved using experimental methods~\cite{pdb2000}. Unfortunately, the prevalent experimental technique, X-ray crystallography, that is used to infer the locations of the atoms of a protein does not provide information about the mechanical -- flexing and bending -- properties of proteins~\cite{kendrew1958}.

Other techniques, both experimental and computational, provide supplementary information about proteins in addition to the information in a PDB file. These include Molecular Dynamics and rigidity analysis, as well as cavity data that describes a protein's surface in 3D.

Molecular Dynamics, MD, is one computational technique developed in the 1970s for inferring the positions and motions of atoms in a protein~\cite{karplus1979}. It involves solving various physics equations to infer how atoms in a protein move in response to repulsion and attraction forces among the amino acids in a protein.

\begin{figure}[h]
    \centering
    \begin{subfigure}[b]{0.22\textwidth}
        \centering
        \includegraphics[width=\textwidth]{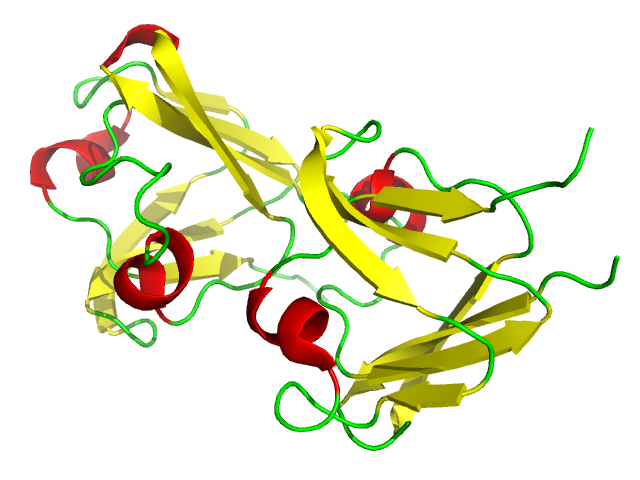}
        \caption{Cartoon}
    \end{subfigure}
    \begin{subfigure}[b]{0.22\textwidth}
        \centering
        \includegraphics[width=\textwidth]{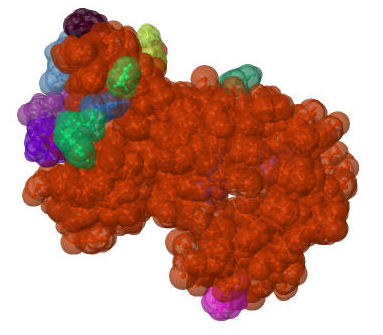}
        \caption{Rigidity Analysis}
    \end{subfigure}\\
    \begin{subfigure}[b]{0.22\textwidth}
        \centering
        \includegraphics[width=\textwidth]{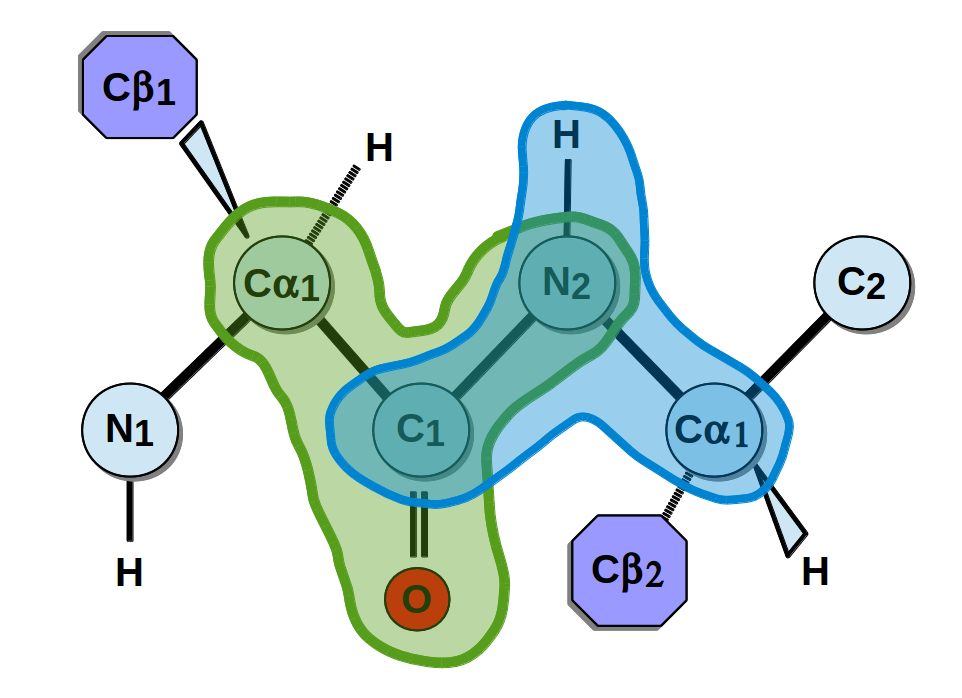}
        \caption{Bodies}
    \end{subfigure}
     \begin{subfigure}[b]{0.18\textwidth}
        \centering
        \includegraphics[width=\textwidth]{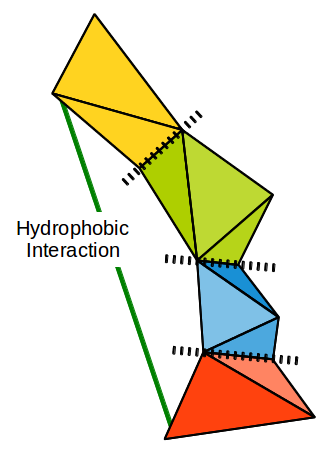}
        \caption{Model}
    \end{subfigure}
    \caption{\small The rigidity properties of biomolecule (PDB structure 4fd9, cartoon rendering in (a)) is shown in (b), where atoms in the same colored region specify atoms that are part of a rigid cluster. Bodies (c) identified from chemical properties of the atoms in the protein are used to generate a mechanical model on which a pebble game algorithm is run to infer the flexible and rigid regions.}
    \label{fig:RigidityAnalysis}
\end{figure}

Rigidity analysis~\cite{jacobs2001,jacobs1995} is a fast graph-based method complimentary to MD, that gives information about a protein's flexibility properties. In rigidity analysis, atoms and their chemical interactions are used to construct a mechanical model of a molecule, in which covalent bonds are represented as hinges, and other stabilizing interactions such as hydrogen bonds and hydrophobics are represented as hinges or bars. A graph is constructed from the mechanical model, and efficient algorithms based on the pebble game paradigm~\cite{jacobs1997} are used to analyze the rigidity of the graph. The rigidity properties of the graph are used to infer the rigidity properties of the protein structure, as shown in Figure~\ref{fig:RigidityAnalysis}.

Information about cavities on a protein's surface is other information that supplements the rigidity and molecular data about a protein structure. There exist efficient algorithms for identifying cavities on the surface of proteins. We use the open source software \textit{fPocket}~\cite{fpocket2010} to identify the cavities and their properties of our approximately 5,000 protein structures. 

Combining these and other pieces of information yields high-dimensional data about each protean structure.

For this work, our data set is built up using biological data from the PDB, rigidity data as output by the KINARI rigidity analysis software~\cite{foxLibrary2012,fox2011}, and cavity data for a protein as output by \textit{fPocket}. For 5,522 protein structures randomly selected from the Protein Databank, we perform rigidity analysis, as well as detected cavities. The combined size of the zipped rigidity, cavity, and PDB data set is 2TB. For each protein structure,  we extract the metrics shown in Table~\ref{tbl:39metrics}.

\begin{table}[h]
  \caption{Summary of 39 dimensional data. Upper-case three-letter combinations refer to the 20 naturally occurring amino acids. }
  \centering
  \begin{tabular}{|p{1.2cm}|p{6.5cm}|}
    \hline
    \textbf{Dimension(s)} & \textbf{Description}  \\
    \hline
    
    1-7 & \textbf{Biological Data, Number of :} atoms, total bonds, hydrogen bonds, single covalent bonds, double covalent bonds, hydrophobic interactions, resonance bonds  \\
    \hline
    8-28 & \textbf{Residue Data, Number of :} total residues, ALA, ARG, ASN, ASP,  CYS, GLN, GLU, GLY, HIS, ILE, LEU, LYS, MET, PHE, PRO, SER, THR, TRP, TYR, VAL \\
    \hline
    29-33 & \textbf{Rigidity Properties, Number of :} Hinges,  Bodies, Degrees of Freedom (DOF), size of largest rigid cluster, average cluster size  \\
    \hline
 34, 35 & \textbf{Largest Cavity :} Surface Area in \AA$^2$, number of residues  \\
 \hline
      36, 37 & \textbf{Second Largest Cavity :} Surface Area in \AA$^2$, number of residues  \\
 \hline
 38, 39 & \textbf{Third Largest Cavity :} Surface Area in \AA$^2$, number of residues  \\   
    \hline
  \end{tabular}
  \label{tbl:39metrics} 
\end{table}

\subsection{Experimental Setup}

Our experimental results are conducted on a 12-core Intel\textsuperscript{\textregistered} Xeon\textsuperscript{\textregistered} 2.00GHz server with a NUMA memory architecture and 16,434,424 kB of memory. 

The number of attributes in our 5,522 protein structures is 39, which gives a theoretical maximum of up to $5522^{28}$ or approximately $6.007576\times 10^{104}$ maximal empty hyper-rectangles per the upper-bounds approach mentioned in \cite{liu1997discovering}. Note that using the approach in \cite{liu1997discovering}, all $5522^{28}$ rectangles would have to be found and then ranked according to size to find the largest ones, which is computationally not feasible.

Our goal is to determine if our algorithm can find the largest rectangles, or, failing that, rectangles that were large enough to be interesting. We conducted 100 experiments with 100,000 rectangles found as a stop condition using expansion strategy 1. Each experiment averaged 20 minutes to complete on one processor and generated approximately 250MB of data. All rectangles along with meta data were output in a plain text file.

\subsection{Experimental Results}

Each of the 100 experiments located the same largest hyper-rectangle, having a volume of  0.4375 on the unit hyper-cube. 

From the hyper-rectangles generated by our algorithm, we wanted to find out which (if any) of the 39 dimensions in our protein data were most frequently the bounding conditions of the largest found hyper-rectangles. 

In addition, we want to infer interesting relationships among the 39 dimensions. For example, is it true that there is a correlation between the number of Alanine (ALA) residues in a protein and the size of the largest cavity? The question posed in the general sense is the following: 

\par
\begingroup
\leftskip2em
\rightskip\leftskip
\noindent
\textbf{If dimension $x$ is a condition that very frequently is a bounding condition of the 100 largest hyper-rectangles, does that coincide with dimension $y$ not being a bounding condition, and vice versa?}
\par
\endgroup

\begin{figure}[h]
    \centering
        \includegraphics[width=0.7\linewidth]{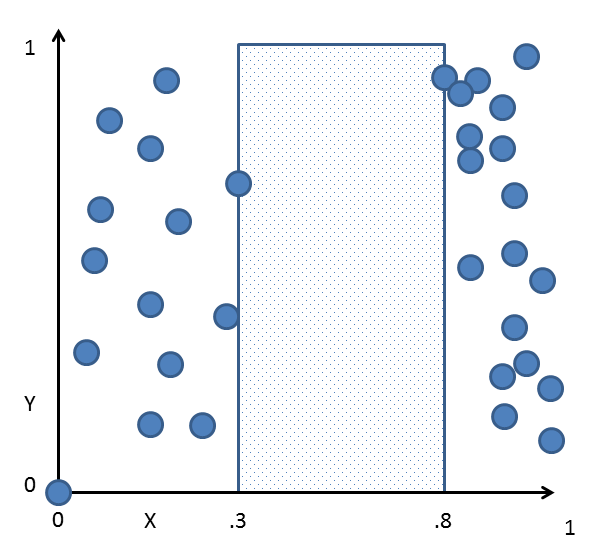}
        \caption{In this 2D example, the maximum rectangle is defined by:  \textit{x} is between 0.3 and 0.8 and \textit{y} is between 0 and 1.} 
    \label{fig:rules}
\end{figure}

To achieve this, we develop scripts to generate if/then rules based on dimensions in our data set that were bounding regions for the largest hyper-rectangles found. The if/then rules are easily explained with a 2D example. In Figure~\ref{fig:rules}, on an input data set specified by the blue points in 2D, the maximal rectangle shown is described by the following rule:

\par
\begingroup
\leftskip2em
\rightskip\leftskip
\noindent
\textbf{\tt If \textit{x} is between 0.3 and 0.8 and \textit{y} is between 0 and 1, then the rectangle bound by these points is empty.}
\par
\endgroup

Extracting the if/then rules from the hyper-rectangle output generated by our algorithm, we find several interesting relationships among the 39-dimensional data.

For instance, whenever we find that Histidine (HIS) count in a protein is in the range 81 to 144, then Glutamic Acid (GLU) is always 34 and the Serine (SER) count is always 74. This implies a relationship between Histidine, Glutamine and Serine counts in cases where the Histidine count is greater than 81 but less than 144. It also implies that Histidine counts specifically, and residue counts in general have the greatest impact on the location of holes within the data set. Although we do not ascribe any biological implications based on these findings, what is important is that these relations would not have been found had we used any  algorithm that had to exhaustively enumerate all large hyper-rectangles.

To investigate the likelihood that our Monte Carlo approach finds the largest hyper-rectangles without needing to enumerate them all, we also explore the number of rectangles that our algorithm considers before finding the (most-likely) largest one. These experiments all involve expansion strategy 1. We observe the following:

\begin{itemize}
\item The least number of rectangles that had to be considered before the largest was found is 117 
\item The highest number of rectangles that had to be considered before the largest was found is 12,183
\item The median number of rectangles that had to be considered before the largest was found is 1,816.5
\item The average number of rectangles that had to be considered before the largest was found is 2724.34
\end{itemize}

From these observations we can deduce that our algorithm only needs to examine a small fraction of the theoretical maximum of $6.007576\times 10^{104}$ possible hyper-rectangles.

\section{Conclusions and future work} \label{conclusions}

To the best of our knowledge, this is the first algorithm for finding holes in high dimensional data that runs in polynomial time with respect to the number of dimensions. Such large hyper-rectangles have been shown to be particularly useful in data mining and data analytics\cite{ordonez1999clustering, liu1998using, edmonds2003mining}. From the MEHRs that we find, we can extract the corresponding if/then rules, which can be used to locate interesting relationships among dimensions. This information can be used for further analysis. 

Our algorithm is a Monte Carlo approach and is amenable for the analysis of high-dimensional data sets. In contrast, existing approaches for finding high-dimensional rectangles are intractable in dimensions 5 or higher because their complexity is exponential in the size of the input, the dimension, or both. We have demonstrated the use of our algorithm on a 39-dimensional data set containing more than 5,000 entries, and we have inferred interesting relationships among the dimensions that could not have been found otherwise.

Why does this approach work so well? Our explanation is that our approach exploits the fact that large hyper-rectangles take up more space and thus are more likely to be found by chance. 

This supports the thesis that a Monte Carlo approach is likely to find large, useful empty hyper-rectangles quickly, simply because the probability of landing in a larger rectangle is greater than landing in a smaller one and in many cases it may be unnecessary to enumerate every hyper-rectangle before concluding that one has likely found the largest.  

To the best of our knowledge there is no publicly available tool, for purchase or otherwise, to analyze holes in data sets.  Because knowledge of these holes has been shown to be useful for both scientific and business purposes~\cite{edmonds2001mining}, we plan to make freely available tools to allow researchers to analyze holes in their own data sets using our method. 

We aim to investigate the overlap between rectangles produced by our expansion strategies. What the overlap implies about the relationship(s) among the dimensions can also be explored.

\newpage


\IEEEtriggeratref{15}



%

\bibliographystyle{ieeetr}
\bibliography{holebbl,jagodzinski}

\end{document}